
\documentclass[twocolumn]{aastex631}

\usepackage[normalem]{ulem}
\usepackage{graphicx}% Include figure files
\usepackage{dcolumn}% Align table columns on decimal point
\usepackage{bm}% bold math

\usepackage{color}
\usepackage{arydshln} 
\usepackage{booktabs}  
\usepackage{threeparttable}
\usepackage{amsmath, amsthm, amssymb}
\usepackage{multirow}
\usepackage{physics}
\usepackage{cuted}
\usepackage{hyperref}
\usepackage{pstricks}
\usepackage{dsfont}

\newcommand{\dr}[1]{\frac{\mathrm{d}#1}{\mathrm{d}r}}
\newcommand{\al}[1]{\textcolor{olive}{#1}}

\shorttitle{Topological mode in stellar oscillations}
\shortauthors{Leclerc et al.}
\begin{document}

\title{Topological modes in stellar oscillations}

\author[0000-0001-6136-4164]{Armand Leclerc}
\affiliation{Univ Lyon, Univ Lyon1, ENS de Lyon, CNRS, Centre de Recherche Astrophysique de Lyon\\ UMR5574, F-69230, Saint-Genis-Laval, France}
\author{Guillaume Laibe}
\affiliation{Univ Lyon, Univ Lyon1, ENS de Lyon, CNRS, Centre de Recherche Astrophysique de Lyon\\ UMR5574, F-69230, Saint-Genis-Laval, France}
\affiliation{Institut Universitaire de France}

\author{Pierre Delplace}
\affiliation{ENS de Lyon,  CNRS, Laboratoire de Physique (UMR CNRS 5672), F-69342 Lyon, France}
\author{Antoine Venaille}
\affiliation{ENS de Lyon,  CNRS, Laboratoire de Physique (UMR CNRS 5672), F-69342 Lyon, France}
\author{Nicolas Perez}
\affiliation{ENS de Lyon,  CNRS, Laboratoire de Physique (UMR CNRS 5672), F-69342 Lyon, France}

\correspondingauthor{Guillaume Laibe}
\email{guillaume.laibe@ens-lyon.fr}



\begin{abstract}
Stellar oscillations can be of topological origin. We reveal this deep and so-far hidden property of stars by establishing a novel parallel between stars and topological insulators. We construct an hermitian problem to derive the expression of the stellar \textit{acoustic-buoyant frequency} $S$ of non-radial adiabatic pulsations. A topological analysis then connects the changes of sign of the acoustic-buoyant frequency to the existence of Lamb-like waves within the star. These topological modes cross the frequency gap and behave as gravity modes at low harmonic degree $\ell$ and as pressure modes at high $\ell$. $S$ is found to change sign at least once in the bulk of most stellar objects, making topological modes ubiquitous across the Hertzsprung-Russel diagram. Some topological modes are also expected to be trapped in regions where the internal structure varies strongly locally.

\end{abstract}



\section{Introduction}

Stars are opaque. Fortunately, deformations of the stellar surface depend on their interiors \cite{Cowling1941,Gough1993,Ledoux1958,Unno1979,CD1996,Aerts2010} and as such, asteroseismology is the Rosetta Stone to infer details of stellar structures \cite{CD1996,Aerts2010}. Stellar spectra consist principally of low-frequencies gravity (g-) modes and high-frequencies {pressure} (p-) modes, defining two bands separated by a finite interval of frequencies, also referred as a gap. The stellar spectrum may also be enriched by additional branches, such as surface wave modes confined in the outer regions. In recent years, a novel type of waves propagating in stratified compressible fluids has been discovered. This so-called \textit{Lamb-like} wave fills the gap between the p- and the g- band. Although this mode bears similarities with the Lamb wave \cite{Lamb1911,Iga2001}, it is confined around peculiar values specific of the stratification profile, and not at the boundaries. The key point is that these waves have been postulated using arguments from topology \cite{Perrot2019}. Modes in the original spatially homogeneous system can be predicted from the analysis of the topological invariant of a simpler dual wave problem with constant coefficients. Similar topological approaches were developed in condensed matter since the eighties, and flourished across all field of physics, including fluid dynamics and plasma over the last few years \cite{hasan2010colloquium,delplace2017topological,shankar2020topological,parker2020topological}.

The Lamb wave has been detected in the atmosphere, but the Lamb-like wave is hardly expected to propagate {on Earth, neither in the atmosphere nor in oceans. Stars were speculated to provide favourable conditions for it to propagate \cite{Perrot2019}. However, this study lacked the treatment of self-gravity, spherical geometry and variations of sound speed, three critical processes as we shall show.} We therefore adapt tools that have been originally developed by the topological insulator community to study the seminal case of adiabatic perturbations of a non-rotating, non-magnetic, stably stratified stellar fluid neglecting gravity perturbations (Cowling's approximation \cite{Cowling1941}). The physical quantities are first rescaled to express the evolution of linear perturbations under the form of a Schrödinger-like wave equation 
\begin{equation}
i\partial_{t}{\mathbf{Y}} = {\mathcal{H}}{\mathbf{Y}} 
\label{eq:schro5x5}
\end{equation}
where $\mathcal{H}=$ \\
\hspace{-0.5cm}$i$\scalebox{0.8}{
$\begin{pmatrix}
0 & 0 & 0 & 0 & -\frac{c_{\rm s}}{r}\partial_\theta\\
0 & 0 & 0 & 0 & -\frac{c_{\rm s}}{r\sin(\theta)}\partial_\phi\\
0 & 0 & 0 & -N & S - c_{\rm s}\partial_r - \frac{c_{\rm s}^\prime}{2}\\
0 & 0 & N & 0 & 0\\
-\frac{c_{\rm s}}{r\sin(\theta)}\partial_\theta(\sin(\theta)\cdot) & -\frac{c_{\rm s}}{r\sin(\theta)}\partial_\phi & -S - c_{\rm s}\partial_r - \frac{c_{\rm s}^\prime}{2} & 0 & 0
\end{pmatrix},$}\\

and the perturbation vector contains rescaled velocities, density and pressure
\begin{equation}
\mathbf{Y} = ^{\top}\!\! \left( \Tilde{u}, \Tilde{v}, \Tilde{w}, \Tilde{\Theta}, \Tilde{p} \right).
\end{equation}
See Appendix~\ref{app:schro} for details.

\noindent As such, the $5\times5$ wave operator $\mathcal{H}$ of the problem is explicitely Hermitian. $\mathcal{H}$  depends on the sound speed $c_\mathrm{s}$, the Brunt-Väisälä frequency $N$ and a characteristic frequency further referred \al{to} as the {\it acoustic-buoyant frequency} $S$ that emerges explicitly
\begin{equation}
S \equiv \frac{c_\mathrm{s}}{2g}\left( N^2 - \frac{g^2}{c_\mathrm{s}^2} \right) - \frac{1}{2} \frac{{\rm d} c_{\rm s}}{{\rm d} r} + \frac{c_\mathrm{s}}{r}.
 \label{eq:S}
\end{equation}
All three parameters vary with radius $r$. Usually, these equations are combined into a single differential equation of high order. Instead, preserving the vectorial structure of the problem is better suited for a topological analysis.

\section{Acoustic-buoyant frequency $S$}

\begin{figure}
    \centering
    \includegraphics{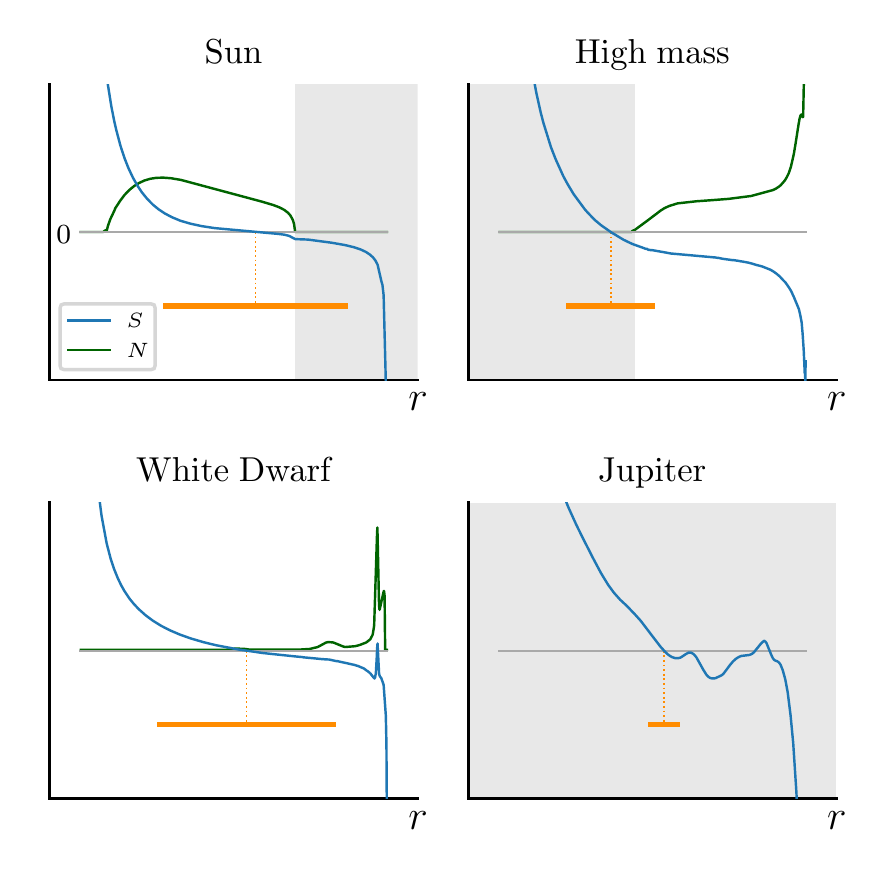}
    \caption{{Profiles of $S$ for four different typical stellar objects. $N$ is plotted for comparison. Solid orange line indicates the region where the topological mode is trapped, as measured by the trapping length $\mathcal{L}$ defined by Eq.~\eqref{eq:trappingLength}. Stellar interiors are computed with \textsc{Mesa}. The High mass star is an $M=100M_\odot$ main-sequence star. The White Dwarf mass is $0.6M_\odot$, during its cooling phase. The Jupiter model has a solid core of 10 Earth masses. $S$ cancels always at least once, whether in the radiative or convective region. Light grey area indicates the convective zone.}}
    \label{fig:HR}
\end{figure}

The acoustic-buoyant frequency $S$ is a coupling parameter for momentum exchange between  buoyant and acoustic oscillations, and was called {\it stratification parameter} in \cite{Perrot2019}. {This role of mode coupling is shown in details below.} Two extra terms appear compared to the plan-parallel case \cite{Perrot2019}: $c_\mathrm{ s} / r$, which accounts for sphericity effects at small radii, and $\frac{1}{2}\dr{c_\mathrm{s}}$, which becomes important when the internal structure of the object varies strongly. $S$ combines the four physical processes responsible for mirror-symmetry breaking in the radial direction: gravity, density stratification, curvature, and radial variations of sound speed. The profile $S(r)$ varies between stellar objects ; however the sound speed is expected to go to 0 at the surface as a positive power law of the density \cite{Chandra1939,Horedt1987}. $S$ is then $-\infty$ at the surface. At small radii, the curvature term guarantees $S$ to reach $+\infty$. $S(r)$ being continuous, it must change sign in the bulk of the star at least once. We confirm this analytically on a stellar polytrope in Appendix~\ref{app:polytrope} and numerically on models of typical stellar objects computed with the \texttt{MESA} code \cite{mesa} (Fig.~\ref{fig:HR}). %$S$ is almost never negligible compared to $N$.%

\begin{figure}[h!]
    \centering
    \includegraphics[width=\columnwidth]{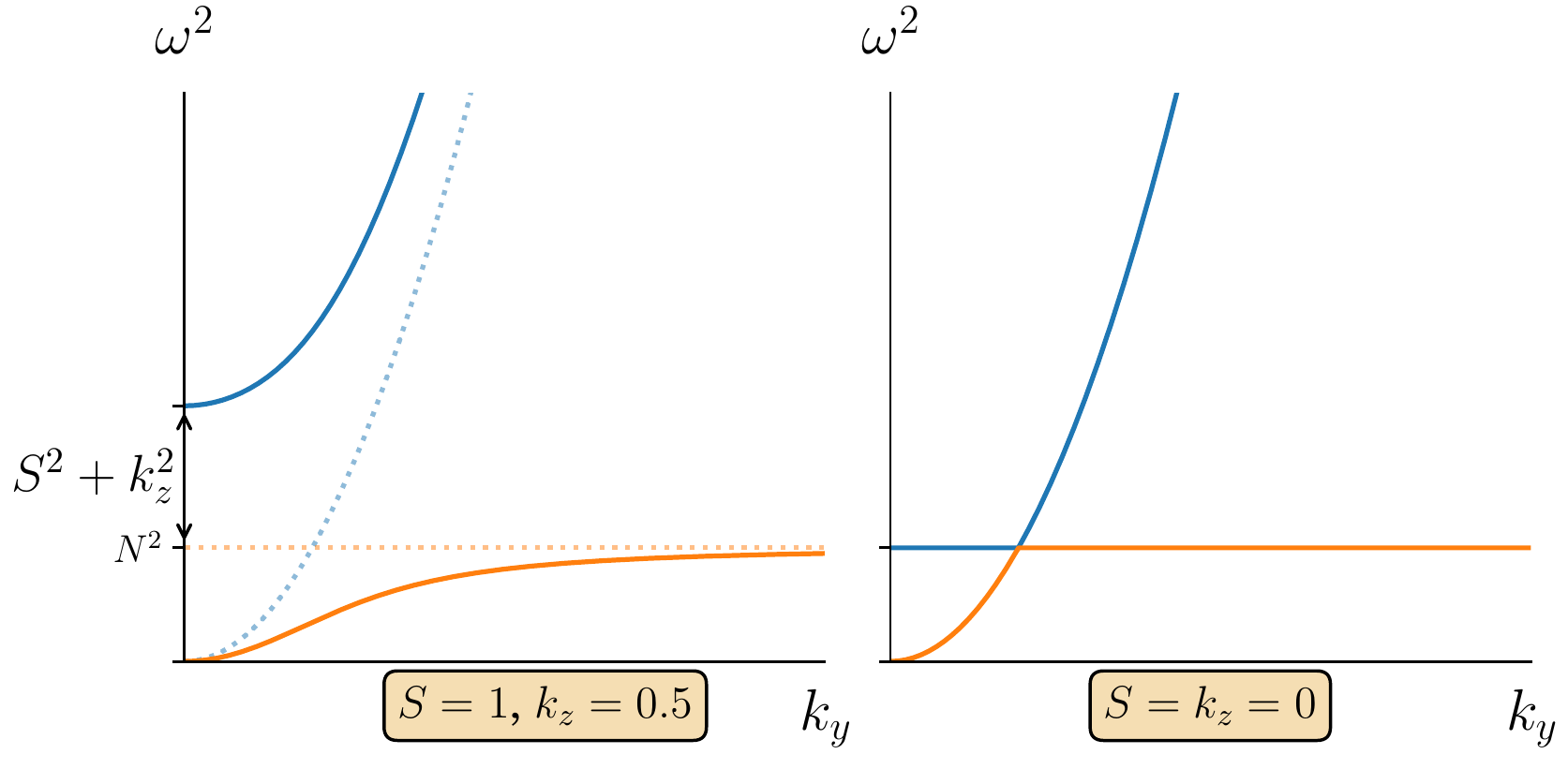}
    \caption{Local dispersion relation of the problem, as modelled by Eqs.~\eqref{eq:coupledWaves1}-\eqref{eq:coupledWaves2}. The p-mode and the g-mode (solid lines) result from the coupling of acoustic and buoyant oscillations (dashed lines for $k_z =0 $). Both $S$ and $k_{z}$ pull the bands away. For any mode, including $k_z = 0$, a gap exists as soon as $S \neq 0$.}
    \label{fig:local_gap}
\end{figure}

The physical nature of the {acoustic-buoyant} frequency $S$ is disclosed by considering the equivalent of Eq.~\eqref{eq:schro5x5} in the 2D plane-parallel $(y,z)$ geometry. After performing a Fourier transform in time and space in the invariant direction $y$ and performing the rescaling $(u,w,\Theta,p) \mapsto c_{\rm s}^{1/2}(u,w,\Theta,p)$, one obtains
\begin{eqnarray}
\partial_t u &=& i{c_\mathrm{s}k_y}p, \label{eqApp:sysReduit1}\\
\partial_t\Theta &=& Nw,\label{eqApp:sysReduit2}\\
\partial_t w &=& -N \Theta - c_{\rm s}\partial_z p + S p,\label{eqApp:sysReduit3}\\
\partial_t p &=& ic_\mathrm{s}k_y u - c_{\rm s}\partial_z w - S w.\label{eqApp:sysReduit4}
\end{eqnarray}
Combining the equations gives
\begin{eqnarray}
(\partial_{tt} + N^2)w = -\partial_t(c_{\rm s} \partial_{z}p - S p),\label{eq:coupledWaves1}\\
(\partial_{tt} + c_\mathrm{s}^2 k_y^2)p = -\partial_t(c_{\rm s} \partial_{z}w + S w),\label{eq:coupledWaves2}
\end{eqnarray}
a system where acoustic and buoyant vibrations are explicitly coupled (no Boussinesq or anelastic approximation is assumed). The first term of the right-hand side of Eq.~\eqref{eq:coupledWaves1} consists of local pressure forces that competes with buoyancy. The first term of the right-hand side of Eq.~\eqref{eq:coupledWaves2} comes from fluid compression in the direction $z$ and is generic from 2D purely acoustic waves. In the long wavelength limit in the stratification direction $z$, these two terms become negligible and
\begin{eqnarray}
(\partial_{tt} + N^2)w = S\partial_t p,\label{eq:Siscoupling_1} \\
(\partial_{tt} + c_\mathrm{s}^2 k_y^2)p = -S\partial_t w , \label{eq:Siscoupling_2}
\end{eqnarray}
showing that $S$ is the frequency of periodic exchanges of momentum between acoustic and buoyant vibrations. Non-Boussinesq contributions allow local densities to be affected by acoustic compression, providing an effect that compets with buoyancy when $S$ is large. Conversely, pressure increases not only through compression, but also through advection in a differential background. These two effects on coupling between g-modes and p-modes were identified by \cite{Lighthill78}. Multiplying Eq.~\eqref{eq:Siscoupling_1} by $\partial_{t}p$ and Eq.~\eqref{eq:Siscoupling_2} by $\partial_{t}w$ shows that the power transmitted by one mode to the other occurs without losses, as expected from the adiabatic assumption. Such a coupling has been widely studied in polariton physics, and shown to result in gap opening \cite{Lagoudakis}. The condition $S = 0$ is therefore associated to local mode decoupling (see Fig.~\ref{fig:local_gap}).

\section{Topological properties of the problem}

Eigenvalues of $\mathcal{H}$ are constrained by topology when varying the physical parameters. These constraints can be efficiently studied by associating a simple matrix to $\mathcal{H}$  that retains the topological constraints. The correspondence is established via a Wigner transform, which allows us to define rigorously a wave that is locally plane without any hypothesis of scale separation (Appendix \ref{app:wigner}). Here, topological properties of $\mathcal{H}$ can be characterised through the eigenvalue problem of the matrix
\begin{eqnarray}
    M &\equiv & 
    \begin{pmatrix}
     N^2 & -NS - iNc_\mathrm{s}k_r\\
     -NS + i N c_\mathrm{s}k_r & L_\ell^2 + S^2 + c_\mathrm{s}^2 k_r^2
    \end{pmatrix},\\
    \omega \mathbf{X} &=& M \mathbf{X} ,\
\end{eqnarray}
where the Lamb frequency is $L_\ell \equiv c_\mathrm{s} \sqrt{\ell(\ell+1)}/r$. $M$ is Hermitian and parametrised by a radial wavenumber $k_{r}$ and parameters $L_\ell$, $c_\mathrm{s}$, $N$ and $S$ that are constant.

\begin{figure*}[t]
\includegraphics[width=\textwidth]{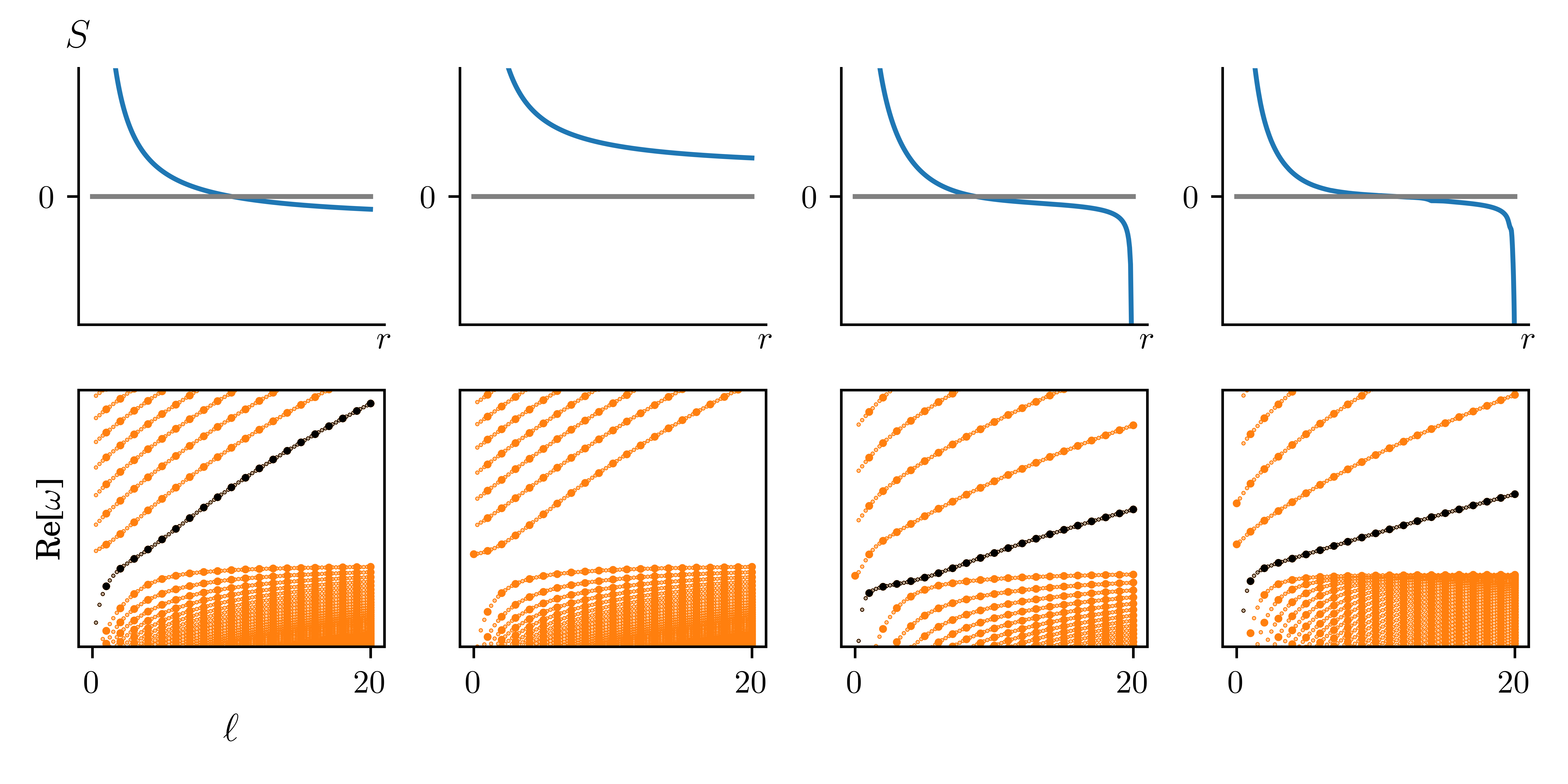}
\caption{A mode develops between the gravity band and the pressure band (bottom) when the {acoustic-buoyant} frequency $S$ (top) of Eq.~\eqref{eq:S} changes sign. From left to right: toy profile that cancels in the bulk, decaying positive profile of $S$, polytrope with polytropic index $n = 3$ and a \textsc{Mesa} Solar-like profile. Physical values of the harmonic degree $\ell$ are integer, and plotted with large points, from 0 to 20. Non-integer values are plotted with small points for readability. Surface gravity waves are filtered out by appropriate boundary conditions. The mode transiting between bands is highlighted in black. These values are computed by solving Eq.~\eqref{eq:schro5x5} numerically using \texttt{Dedalus} \cite{dedalus}.}
\label{fig:spectra}
\end{figure*}

As expected, the two eigenvalues of $M$ correspond to the square of the frequencies of the local pressure and gravity modes. Interestingly, these two bands intersect when $k_r=0$, $L_\ell=N$, $S=0$ for any value of $c_\mathrm{s}$ and $N$, i.e. the two frequencies degenerate into a single one (see Appendix~\ref{app:wigner}). Such a degeneracy point behaves like a topological monopole in parameter space ($k_r,L_\ell,S$), which is {characterised by} an integer called the Chern number \cite{Chern1946}. A non-zero Chern number translates the topological obstruction to smoothly define the phase of the eigenvectors   -- that describe the local polarization relations of $M$ -- all around the degeneracy point in parameter space. In that case, the eigenvectors can only be defined smoothly over patches in parameter space, corresponding to different gauge choices. The $U(1)$ gauge transformation that connects the different patches is a phase whose winding is the Chern number. In our case, we find the Chern numbers associated to  the gravity and the pressure bands to be $\mathcal{C}^g = +1$ and $\mathcal{C}^p = -1$ respectively (see Appendix~\ref{app:chern} for computations).
These topological considerations can be back-connected to the original problem : any change of sign of the {acoustic-buoyant} frequency $S(r)$ is associated to the existence of a branch that transits from the g-band towards the p-band as $\ell$ increases. Mathematically, this correspondence is ensured by index theorems \cite{Atiyah1963,Chern1946,Perrot2019,Faure2019,notesPiR,Nakahara,esposito97}. {The transiting branch flows from the upper-band to the lower-band or \textit{vice-versa}, depending on the sign of $S'$ at the change of sign of $S$. In stars, $S'<0$ and the mode transits from the g- to the p- band: this mode is the Lamb-like wave} \cite{Perrot2019}. Fig.~\ref{fig:spectra} confirms the deep relation between a change of sign of $S(r)$ and the existence of a mode transiting from the g-band at small $\ell$ to the p-band at large $\ell$. {The physical validity of this mode is carefully verified in Appendix~\ref{app:CL}.}

By analogy with similar modes encountered in a variety of other physical systems \cite{hasan2010colloquium,delplace2017topological,shankar2020topological,parker2020topological}, one may expect for the global stellar mode to have no node, and to transit between the bands at a value of $\ell$ such that $L_\ell \sim N$. One may also expect for the eigenfunctions to be located around the radius $r_0$ where $S(r_0) = 0$. These properties of the topological mode can be verified on a simple analytically solvable model presented in the next section.

\section{Topological mode in analytical model}

We present a simple analytical model featuring a cancellation in $S$, and show that the analytical solution of the wave equation include the topological mode. Consider a fluid where all quantities but $S$ are constant in space:

\begin{eqnarray}
    S(r) &=& -\alpha(r-r_0),\label{eq:Slinear}\\
    N(r) &=& N_0,\\
    c_\mathrm{s}(r) &=& c_\mathrm{s,0},\\
    L_\ell^2(r) &=& L_{\ell,0}^2 = c_\mathrm{s,0}\frac{\ell(\ell+1)}{r_0^2}.
\end{eqnarray}
This parametrisation mimics a situation where variations of $S$ would be infinitely more abrupt than the other quantities. {In this minimal model, $S$ vary linearly and cancels in $r_0$. This model} may {thus be} thought of as the compressible-stratified analogue to the equatorial shallow water model solved by Matsuno \cite{Matsuno1966}. Perform the transform $(u,v,w,\Theta,p) \mapsto c_{\rm s}^{1/2}(u,v,w,\Theta,p)$ in
Eq.~\eqref{eq:schro5x5}, then apply a time-Fourier Transform and project on spherical harmonics. The variables combine into a a single ODE on $p$
\begin{equation}
    \left(\frac{\mathrm{d}^2}{\mathrm{d}r^2} -\left(\frac{S(r)}{c_\mathrm{s,0}}\right)^2 - \left(\frac{S(r)}{c_\mathrm{s,0}}\right)^\prime + k_{r,0}^2 \right)p = 0,
    \label{eq:schroWell}
\end{equation}
where we denote
\begin{equation}
    k_{r,0}^2 \equiv \frac{(L_{\ell,0}^2-\omega^2)(N_0^2 - \omega^2)}{c_\mathrm{s,0}^2\omega^2},
\end{equation}
and use the symbol ${}^\prime$ for derivatives with respect to $r$ for background quantities. Eq.~\eqref{eq:schroWell} holds for any $S(r)$, and can be seen as a Schrödinger equation describing a particle of energy $k_{r,0}^2$ in the potential $V = S^2/c_\mathrm{s,0}^2 + S^\prime/c_\mathrm{s,0}$. For the model of Eq.~\eqref{eq:Slinear}, this reduces to
\begin{equation}
\left[ \frac{\mathrm{d}^2}{\mathrm{d}x^2} - \left(\frac{1}{4}x^2 - \frac{1}{2}(1+\frac{c_\mathrm{s,0}}{\alpha} k_{r,0}^2) \right) \right]p = 0 ,
\label{eq:hermite}
\end{equation}
using the dimensionless quantity $x \equiv \sqrt{2}\sqrt{\frac{\alpha}{c_\mathrm{s,0}}}(r-r_0)$. The solution is a Parabolic Cylinder Function $U$ \cite{abramowitz1972}
\begin{equation}
    p = U\left(-\frac{1}{2}(1+\frac{c_\mathrm{s,0}}{\alpha}k_{r,0}^2),~ x\right).
    \label{eq:paraCylFunc}
\end{equation} Regularity at infinity imposes the first argument to be negative half-integer, leading to the quantization
\begin{equation}
\frac{c_\mathrm{s,0}}{\alpha}k_{r,0}^2 = 2n+1,
\label{eq:quantifHermite}
\end{equation}
for any $n \in \mathds{N}$. 

\begin{figure}[h!]
\includegraphics[width=\columnwidth]{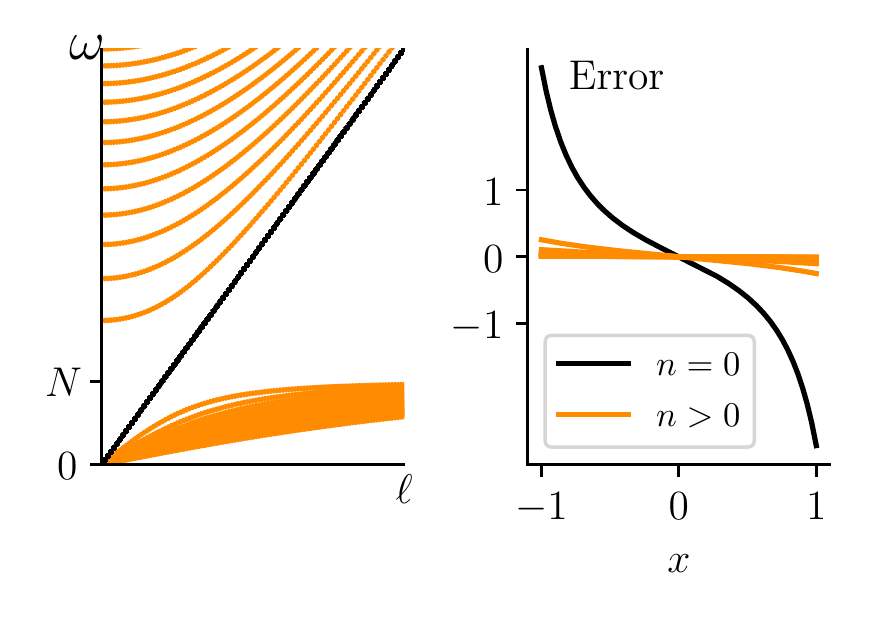}
\caption{Left: spectrum of the minimal model parametrized by Eq.~\eqref{eq:Slinear}. The topological mode is the $n=0$ mode, and transits between the bands. Right: Measure of the error a JWKB approximation of the solutions would make. The error on the $n=0$ mode is  not small.}
\label{fig:spectre_hermite}
\end{figure} 

Solutions reduce then to Hermite functions ${p = {\rm e}^{-x^2/2}H_n(x)}$, where $H_n$ denotes the $n$th Hermite polynomial. Fig.~\ref{fig:spectre_hermite} shows the spectrum associated to this problem. The values of $\omega$ can be inverted in Eq.~\eqref{eq:quantifHermite}. For $n \geqslant 1$, each value of $k_{r,0}$ give two eigenfrequencies, the $n$th g-mode and the $n$th p-mode. The expected topological mode corresponds to $n = 0$. One of the two eigenfrequencies associated with this solution is unphysical, since the eigenfunctions diverges quickly at infinity. The other verifies
\begin{equation}
    \omega = L_{\ell,0},
\end{equation}
which transits between the bands as $\ell$ increases, as shown on Fig.~\ref{fig:spectre_hermite}. This property is associated to the fact that ${S'(r_0) = -\alpha< 0}$ at the cancellation point.\\
The topological mode has the profile
\begin{equation}
    p(r) = p_0 \exp\left( - \frac{\alpha}{c_\mathrm{s,0}}(r-r_0)^2 \right),
\end{equation}
an expression that provides a definition of the length over which the mode has significant amplitude
\begin{equation}
    \mathcal{L} \equiv \sqrt{\frac{c_\mathrm{s,0}}{\alpha}} = \sqrt{{c_\mathrm{s,0}}/{\Big|\dr{S}\Big|_{S=0}}} ,\label{eq:trappingLength}
\end{equation}
which we call the {\it trapping length}. Denoting ${R(x)\equiv-\frac{1}{4}x^2+n+1/2}$ the second term of Eq.~\eqref{eq:hermite} that corresponds to a solution for a given $n$, we find for JWKB approximation of the solution to be valid when the condition $|R^{-3/2}\frac{\mathrm{d}R}{\mathrm{d}x}| \ll 1$ is satisfied \citep{daghigh12}. Fig.~\ref{fig:spectre_hermite} shows this quantity for the first modes. The topological mode $n=0$ breaks strongly this validity condition. As expected, JWKB techniques cannot capture the topological mode.

This analytical solution confirms that the topological mode is the mode with zero node of the system, and that this mode is not accessible with scale separation methods.

\section{Discussion}

\begin{figure}[h!]
    \centering
    \includegraphics[width=\columnwidth]{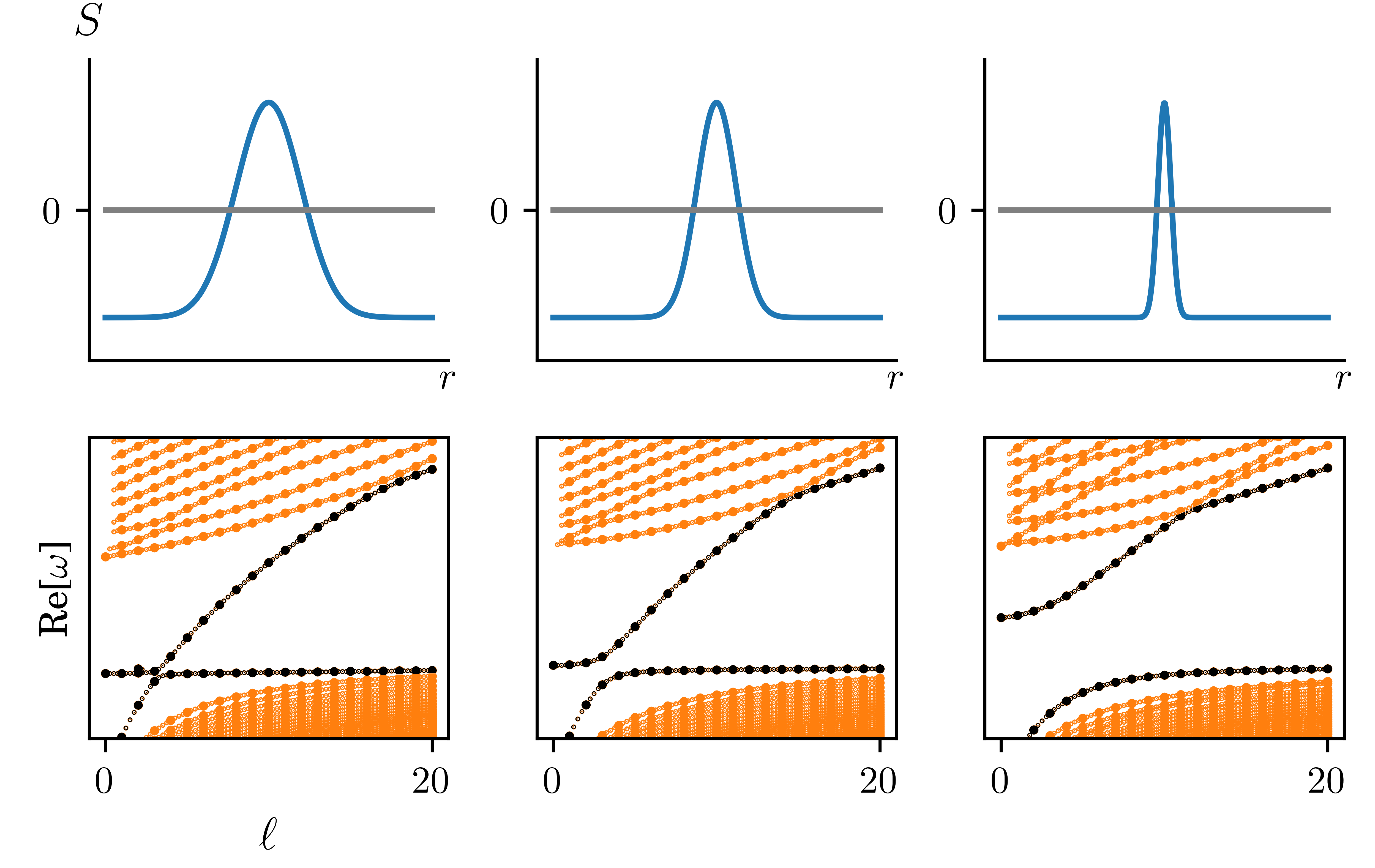}
    \caption{Peaks of $S$ through positive values imply the existence of two topological modes in the spectrum. The sharper the peak, the more the modes hybridize, and their branches avoid crossing in the spectrum.}
    \label{fig:type2}
\end{figure}

Interestingly, the topological mode and the surface-gravity mode have both zero node and similar dispersion relations. Numerical experiments show that when they coexist, they hybridize to form a unique $n=0$ mode. {A comprehensive study including various boundary conditions is performed in Appendix~\ref{app:fmode}}. We interpret this hybridised mode as the {\it f-mode} of asteroseismology \cite{Gough1993,Rozelot2011}, revealing its previously unexpected hybrid nature. \\
{Finally, strong local gradients of thermodynamical quantities may give rise to peaks of acoustic-buoyant frequency where $S$ changes sign twice over a short scale, as in the White Dwarf model showed on Fig.~\ref{fig:HR}. This results in two modes of topological origin that may be used to probe fine details of the structure of the stellar object. The white dwarf is the canonical object for application of this study, as it is fully radiative. Its profile of $S$ cancels three times, two of them resulting from a phase transition close to the surface. For this model, we predict three topological modes, one for each cancellation. One crossing the gap, with long trapping length $\mathcal{L}$, as the slope of $S$ where it changes sign is low at the first cancellation. Two more modes with zero nodes are predicted close to the peak of $S$ just underneath the surface, with much smaller trapping lengths $\mathcal{L}$, as the slope of $S$ is high when $S$ changes sign. They potentially overlap each other, such that they would hybridise. This hybridisation could serve as a measure of the peak in $S$, meaning the modes could serve as probes for the associated phase transition. This hybridization is illustrated on Fig.~\ref{fig:type2}.}\\
The current study focuses on stably stratified stars, for which index theorems on Hermitian systems apply. However the effect of a convective zone on the Lamb-like wave remains to be investigated. Such a region, where $N^2$ vanishes, is indeed sustained by the convective circulation of the background. Fig.~\ref{fig:HR} shows that in the Sun, $S$ cancels in the radiative zone, close to the convective zone. The trapping length of the topological mode indicates interactions with the convective zone, although convection is out of the scope of this study. In High Mass stars, $S$ cancels within the convective core where the topological mode is not guaranteed by this study, as no background flow is considered in the wave equation Eq.~\eqref{eq:schro5x5}. The same conclusion applies to Jupiter, which is fully convective, and has interesting multiple cancellations of $S$.

{Lamb-like waves are neither Lamb waves, surface-gravity waves, nor mixed modes \cite{Dziembowski2001,Dupret2009,Deheuvels2010}. Mixed modes are linear combinations of g-modes and p-modes standing in different cavities in the star, due to spatial variations of $N$ and $L_\ell$ and can have a high number of nodes. The Lamb-like wave emanates as a mode $n=0$ of a single cavity hosting both g-modes and p-modes.}\\
We expect generic properties of stellar pulsations related to topology such as ray tracing to be encoded in $S(r)$ \cite{Perez2021}. Other discrete symmetries can be broken in the presence of rotation \cite{Perez2021b} and magnetic fields \cite{Cally2006,Parker2020}, and one should expect the emergence of new classes of topological waves when these additional ingredients are taken into account, potentially at the stellar tachocline where strong shear develops. The resilience of these topological modes on unstable stratification when $N^{2} < 0$, or with the inclusion of dissipative effects, is a highly promising avenue of research in the currently flourishing field of non-Hermitian topological waves \cite{Delplace2021,Gong2018,Yao2018,Bergholtz2021}.  

\section{Conclusion}

In this study, we revisit the old field of stellar pulsations under the bright new prism of topology. By doing a novel parallel between stars and topological insulators, we establish for the first time the existence of a wave of topological origin in stars. We derive the expression of a novel key physical parameter, the acoustic-buoyant frequency. We demonstrate in a comprehensive analysis that topological modes are associated to zeros of this frequency, and show the ubiquitous existence of at least one topological mode across the entire spectrum of stellar object in the Universe. More importantly, we show that local phase transitions, which are key for understanding the evolution of stars within the cosmological context, may give rise to pairs of robust topological modes. The hunt of these modes may therefore become a critical target for future cutting-edge instruments such as the \textit{PLATO} mission.

%%%%%%%%%%%%%%%%%%%%%%%%%%%%%%%%%%%%%%%%%%%%%%%%%%%%%%%%%%%%%%%%%%%%%%%%%%%%%%%%
\vspace{2cm}
\section*{Acknowledgments}
A.L. lead the derivation of the {acoustic-buoyant} frequency and performed numerical simulations. G.L. lead the astrophysical analysis and the writing of the manuscript. P.D. and A.V. lead the topological analysis and the analogy with the plane-parallel case. N.P. performed the numerical experiments on surface-gravity wave and topological mode hybridisation.\\
G.L. acknowledges funding from the ERC CoG project PODCAST No 864965. This project has received funding from the European Union's Horizon 2020 research and innovation programme under the Marie Sk\l odowska-Curie grant agreement No 823823. This project was partly supported by the IDEXLyon project (contract nANR-16-IDEX-0005) under the auspices University of Lyon. We acknowledge financial support from the national programs (PNP, PNPS, PCMI) of CNRS/INSU, CEA, and CNES, France. AV and PD were supported by the national grant ANR-18-CE30-0002-01 and Idex Tore. N.P. was funded by a PhD grant allocation Contrat doctoral Normalien. We thank S. Deheuvels, I. Baraffe, G. Chabrier, E. Jaupart, J. Fensch and E. Lynch for useful comments and discussions. The authors are grateful to the anonymous referee, whose thorough comments helped to improved the quality of this article significantly.

\newpage
\appendix

\section{Wave equation as Schrödinger-like}
\label{app:schro}

%\subsection{$5\times5$ system}
We study the evolution of a perturbation in velocity, pressure and density of a stable equilibrium of a star. $(w',u',v')$ is the perturbation's velocity in spherical coordinates, $\rho'$ and $p'$ are the perturbations in density and pressure. The system of equations is obtained by linearizing the equations of mass and momentum conservation assuming adiabatic evolution. As a first step, the hermiticity of the linear system is made explicit by the mean of the following transformation 
\begin{equation}
  \label{eq:transform}
  \begin{aligned}
    (w',u',v') &\mapsto  (\Tilde{w},\Tilde{u},\Tilde{v}) = \rho_0^{1/2}r(w',u',v'),\\        
    p' &\mapsto \Tilde{p} = \rho_0^{-1/2}c_\mathrm{s}^{-1}rp',\\
\rho' &\mapsto \Tilde{\Theta} = \rho_0^{-1/2}r\frac{g}{N}(\rho' - \frac{1}{c_\mathrm{s}^2}p'),
  \end{aligned}
\end{equation}
where $\Tilde{\Theta}$ is the potential density of the fluid. The evolution of the perturbation is then 
\begin{equation}
i\partial_{t}{\mathbf{Y}} = {\mathcal{H}}{\mathbf{Y}},
\label{eqApp:3Dschro}
\end{equation}
where $\mathcal{H}$ is the differential operator\\
$$i\begin{pmatrix}
0 & 0 & 0 & 0 & -\frac{c_{\rm s}}{r}\partial_\theta\\
0 & 0 & 0 & 0 & -\frac{c_{\rm s}}{r\sin(\theta)}\partial_\phi\\
0 & 0 & 0 & -N & S - c_{\rm s}\partial_r - \frac{c_{\rm s}^\prime}{2}\\
0 & 0 & N & 0 & 0\\
-\frac{c_{\rm s}}{r\sin(\theta)}\partial_\theta(\sin(\theta)\cdot) & -\frac{c_{\rm s}}{r\sin(\theta)}\partial_\phi & -S - c_{\rm s}\partial_r - \frac{c_{\rm s}^\prime}{2} & 0 & 0
\end{pmatrix} ,
$$

the perturbation vector is
\begin{equation}
\mathbf{Y} = ^{\top}\!\! \left( \Tilde{u}, \Tilde{v}, \Tilde{w}, \Tilde{\Theta}, \Tilde{p} \right).
\end{equation}
and $c_\mathrm{s}^\prime \equiv \dr{c_\mathrm{s}}$. This rescaled system of equations reveals that three functions govern the perturbations: $c_\mathrm{s}(r)$, $N(r)$ and the {acoustic-buoyant} frequency
\begin{equation}
    S(r) \equiv \frac{c_\mathrm{s}}{2g}\left( N^2 - \frac{g^2}{c_\mathrm{s}^2} \right) - \frac{1}{2} \frac{{\rm d} c_{\rm s}}{{\rm d} r} + \frac{c_\mathrm{s}}{r}.
\end{equation}.

%%%%%%%%%%%%%%%%%%%%%%%%%%%%%%%%%%%%%%%%%%%%%%%%%%%%%%%%%%%%%
\section{Polytropic stars}
\label{app:polytrope}

We derive the expressions for the parameters $c_\mathrm{s}(r),~N(r),~S(r)$ for polytropic stars that verify the equation of state $P = k \rho^{1+1/n}$. {Static equilibrium satisfies the continuity and Poisson equations, and is given by a seminal solution in terms of the Lane-Emden equation}

\begin{eqnarray}
    \frac{1}{x^2}\frac{\mathrm{d}}{\mathrm{d}x}( x^2\frac{\mathrm{d}f}{\mathrm{d}x})+f^n=0,\\
    f(0)=1,\\
    f'(0)=0,
\end{eqnarray}
where $\rho(r) = \rho_c f^n(x=r/a)$, and $a^2 = (n+1)\frac{P_c}{4\pi \mathcal{G}\rho_c^2}$. We adopt length and time units such that $a = 1$ and $k\rho_c^{1/n} = P_c/\rho_c = 1$ and assume the fluid to be a monoatomic perfect gas ($\Gamma_1$ = 5/3). We then obtain

\begin{eqnarray}
    c_\mathrm{s}^2 &=& \frac{5}{3} f,\\
    S &=& \sqrt{\frac{5}{3}} f^{1/2} \left(\frac{1}{r} + \alpha(n)\frac{f'}{f}\right),\\
    N^2 &=& \frac{2}{3}(n+1)\left(n-\frac{3}{2}\right)\frac{f'^2}{f},
\end{eqnarray}
where $\alpha(n) = (n+7/3)/10$. Since $f(r=R) = 0$ and $f'(r=R)<0$, a polytropic star verifies $S(r=R) = -\infty$.

%%%%%%%%%%%%%%%%%%%%%%%%%%%%%%%%%%%%%%%%%%%%%%%%%%%%%%%%%%%%%
\section{Local properties of $\mathcal{H}$: Wigner transform}
\label{app:wigner}
Symbolic calculus gives a way to associate to differential operators acting on functions other functions called {\it symbols} acting on a phase space. The symbol $\text{Symb}\left[\hat{f}\right]$ of an operator $\hat{f}$ (e.g. a differential operator) is obtained by a Wigner transform, defined as
\begin{equation}
\text{Symb}[\hat{f}] \equiv f(x,k) = \int \mathrm{d}y K_{\hat{f}}\left( x + \frac{y}{2}, x - \frac{y}{2} \right)\mathrm{e}^{-i k y},
\end{equation}
where $K_{\hat{f}}$ is the integral kernel of the operator $\hat{f}$: $(\hat{f}\Psi)(x) \equiv \frac{1}{2\pi}\int \mathrm{d}y K_{\hat{f}}\left(x,y\right)\Psi(y)$.\\
The inverse correspondence is the Weyl quantification of the symbol
\begin{eqnarray}
&&\hat{f}=\mathrm{Op}\left[f(x,k)\right] \nonumber\\
&&\equiv \frac{1}{(2\pi)^{2}}\int \mathrm{d}x\mathrm{d}k \, \mathrm{d}\xi\mathrm{d}\eta f(x,k)\mathrm{e}^{i \left[ \xi(x-\hat{x}) + \eta(k-\hat{k})\right]} ,
\label{eqApp:ww_def}
\end{eqnarray}
such that $\mathrm{Symb}\left[\mathrm{Op}\left[f(x,k)\right]\right]=f(x,k)$. Eq.~\ref{eqApp:ww_def} often gives a convenient way to relate an operator to a given functional form of its symbol. For example, the Wigner symbol of the differential operator $\partial_x$ is Symb$\left[\partial_x\right] = - i k_x$, a similar expression as the Fourier transform in this case. Weyl quantification involves commutators $\left[x,\partial_x\right] = -1 \neq 0$, that provide a rigorous framework for making a correspondence of differential operators with varying coefficients to a phase space and this, without assumptions on the wavelengths of its eigenfunctions contrary to JWKB approaches \cite{Faure2019,onuki2020quasi,venaille2020wave}. Hence the following relation: 
\begin{eqnarray}
\mathrm{Op}\left[f(x)k\right] &\equiv & \frac{1}{(2\pi)^{2}}\int \!\!\! \int \!\!\! \int \!\!\! \int \mathrm{d}x\mathrm{d}k \, \mathrm{d}\xi\mathrm{d}\eta \: f(x)k\:\mathrm{e}^{i \left[ \xi(x-\hat{x}) + \eta(k-\hat{k})\right]} , \\
&=& \frac{1}{(2\pi)^{2}}\int \!\!\! \int \!\!\! \int \!\!\! \int \mathrm{d}x\mathrm{d}k \, \mathrm{d}\xi\mathrm{d}\eta \: f(x)k\: \mathrm{e}^{i \xi(x + \eta/2) + i\eta k}\mathrm{e}^{-i\xi\hat{x}}\mathrm{e}^{-i\eta\hat{k}} , \\ 
&=& \frac{1}{(2\pi)^{2}}\int \!\!\! \int \!\!\! \int \!\!\! \int \mathrm{d}x\mathrm{d}k \: \, \mathrm{d}\xi\mathrm{d}\eta f(x-\frac{\eta}{2})k \mathrm{e}^{i \xi x  + i\eta k}\mathrm{e}^{-i\xi\hat{x}}\mathrm{e}^{-i\eta\hat{k}} ,\\
&=& \int \!\!\! \int \mathrm{d}x\mathrm{d}k \:f(x)k\: \delta(x-\hat{x})\delta(k-\hat{k}) \; \\
&&\mspace{70mu}+ \; \int \!\!\! \int \mathrm{d}x\mathrm{d}k \:\frac{i}{2}f^{\prime}(x)k\: \delta(x-\hat{x})\delta^{\prime}(k-\hat{k}) , \\
&=& f(\hat{x}) \hat{k} - \frac{i}{2}f'(\hat{x}). \label{eq:WWcdr}
\end{eqnarray}
Microlocal analysis connects topological properties of the eigenvectors of the Wigner symbol to spectral properties of the operator $\mathcal{H}$. The correspondence relies on index theorems \cite{Atiyah1963}, and provide a powerful tool to identify spectral properties of an operator from a much simpler scalar dual problem. In particular, this procedure allows analysis at long wavelengths that are filtered out by JWKB approximation.\\
The operator $\mathcal{H}$ depends on parameters that vary with radius $r$. Key manipulation concerns the term
\begin{eqnarray}
\mathrm{Symb}\left[\frac{c_\mathrm{s}}{2g}\left(N^2 - \frac{g^2}{c_\mathrm{s}^2} \right) - c_\mathrm{s}^\prime - c_\mathrm{s}\partial_r\right] &=& \mathrm{Symb}\left[\frac{c_\mathrm{s}}{2g}\left(N^2 - \frac{g^2}{c_\mathrm{s}^2} \right) - \frac{c_\mathrm{s}^\prime}{2} - c_\mathrm{s}\partial_r - \frac{c_\mathrm{s}^\prime}{2}\right] \\
&=& \mathrm{Symb}\left[\frac{c_\mathrm{s}}{2g}\left(N^2 - \frac{g^2}{c_\mathrm{s}^2} \right) - \frac{c_\mathrm{s}^\prime}{2}\right] - \mathrm{Symb}\left[c_\mathrm{s}\partial_r + \frac{c_\mathrm{s}^\prime}{2}\right] \\
&=& \frac{c_\mathrm{s}}{2g}\left(N^2 - \frac{g^2}{c_\mathrm{s}^2} \right) - \frac{c_\mathrm{s}^\prime}{2} - ic_\mathrm{s}k_r\\
&=& S - ic_\mathrm{s}k_r,
\end{eqnarray}
applying identity Eq.~\eqref{eq:WWcdr} with $f=c_\mathrm{s}$. The Wigner symbol of $\mathcal{H}$ in the radial direction is then
\begin{equation}
H \equiv i\begin{pmatrix}
0 & 0 & 0 & 0 & -\frac{c_{\rm s}}{r}\partial_\theta\\
0 & 0 & 0 & 0 & -\frac{c_{\rm s}}{r\sin(\theta)}\partial_\phi\\
0 & 0 & 0 & -N & S + ic_\mathrm{s}k_r\\
0 & 0 & N & 0 & 0\\
-\frac{c_{\rm s}}{r\sin(\theta)}\partial_\theta(\sin(\theta)\cdot) & -\frac{c_{\rm s}}{r\sin(\theta)}\partial_\phi & -S + ic_\mathrm{s}k_r & 0 & 0
\end{pmatrix} .
\end{equation}
The object $H $ is a function with respect to $r$, and an operator over the angles $(\theta,\phi)$. Performing a Fourier transform with respect to time, one obtains
$$ 
-\omega \Psi = H\Psi,
$$
which gives
$$
\omega^2 \Psi = H^2 \Psi.
$$
The operator $H^2$ is block diagonal
\begin{equation}
H^2 = 
\begin{pmatrix}
A & 0_{3,2}\\
0_{2,3} & \bar{M}\\
\end{pmatrix},
\end{equation}
where $0_{2,3}$ and $0_{3,2}$ denote null matrices of dimensions $2\times3$ and $3\times2$ respectively. The eigenvalues $\omega^2$ of $H^2$ consist generically of the union of both the eigenvalues of $A$ and $\bar{M}$. Here, the eigenvalues of $A$ and $\bar{M}$ are the same. Indeed, a nonzero eigenvector $\Psi = ^{\top}\!\! \left( \Tilde{u}, \Tilde{v}, \Tilde{w}, \Tilde{\Theta}, \Tilde{p} \right)$ cannot have $\Tilde{\Theta}= \Tilde{p}=0$ or  $\Tilde{u}= \Tilde{v}= \Tilde{w}=0$, as a perturbation cannot be made of only velocity with no pressure/density or pressure/density with no velocity. Hence, no eigenvector of $H^2$ can be of the form $ ^{\top}(u,v,w,0,0)$ or of the form $ ^{\top}(0,0,0,\Theta,p)$, implying that $A$ and $\bar{M}$ cannot have different eigenvalues. The eigenvalues of $H^2$ are therefore the eigenvalues of $\bar{M}$, which is the $2\times2$ matrix
\begin{equation}
    \bar{M} \equiv 
    \begin{pmatrix}
     N^2 & -NS - iNc_\mathrm{s}k_r\\
     -NS + i N c_\mathrm{s}k_r & -\frac{c_\mathrm{s}^2}{r^2}\mathcal{L} + S^2 + c_\mathrm{s}^2 k_r^2
    \end{pmatrix},
\end{equation}
where $\mathcal{L} \equiv \frac{1}{\sin(\theta)}\partial_\theta(\sin(\theta) \partial_\theta\cdot)+\frac{1}{\sin(\theta)^2}\partial_{\phi \phi}$.
After projecting onto spherical harmonics $Y_\ell^m$, one obtains the matrix $M$
\begin{equation}
    M \equiv 
    \begin{pmatrix}
     N^2 & -NS - iNc_\mathrm{s}k_r\\
     -NS + i N c_\mathrm{s}k_r & L_\ell^2 + S^2 + c_\mathrm{s}^2 k_r^2
    \end{pmatrix} ,
\end{equation}
where $L_\ell$ is the Lamb frequency as presented in the main text.

The matrix $M$ is hermitian and as such is diagonalisable. Its two eigenvalues are degenerate and both take the value $N^2$ when
\begin{equation}
    \begin{pmatrix}
     k_r\\
     L_\ell\\
     S
    \end{pmatrix}
    =
    \begin{pmatrix}
     0\\N\\0
    \end{pmatrix}
    \equiv
    \mathbf{p_0}.
\end{equation}
This degenerescence is identical as the one found in \cite{Perrot2019}. An other degenerescence of the eigenvalues occur at ${L_\ell=-N}$, which we will ignore as it corresponds to negative values of $L_\ell$.\\

%%%%%%%%%%%%%%%%%%%%%%%
\section{Chern numbers}
\label{app:chern}
The first Chern number $\mathcal{C}^{\left( n \right)}$ of the $n-$th band is the topological charge associated to the flux of the Berry curvature $\mathbf{F}^{\left( n\right)}$ over a close surface $\Sigma$ of the parameter space of the matrix $M$ \cite{Chern1946}. Its expression is
\begin{equation}
    \mathcal{C}^{\left( n \right)}~=~\frac{1}{2\pi} \int_{\Sigma}~\mathbf{F}^{\left( n\right)}\cdot\mathrm{d} \mathbf{\Sigma} .
    \label{eqApp:chernNum}
\end{equation}
where we denote $n=1$ for the g-band and $n=2$ for the p-band. The relevant parameter space for our study is $\{k_r,L_\ell,S\}$, such that the Berry curvature is a vector with three components denoted $\left(F^{(n)}_{k_r,L_\ell},\; F^{(n)}_{L_\ell,S},\; F^{(n)}_{S,k_r} \right)$. A degeneracy point for the eigenvalues of the symbol matrix $M$ is topologically non-trivial if the Chern numbers $\mathcal{C}^{\left( n \right)}$ take non-zero values at this point. This topological property is reflected in the spectrum of the original operator problem $\mathcal{M}$ by $\abs{\mathcal{C}^{\left( n \right)}}$ modes that transit from one band to another.\\
To calculate the value of the Chern numbers of $M$ at the degeneracy point $\mathbf{p_0}$, we start from the definition
\begin{equation}
    F_{p,p'} = i \left(\partial_p \Psi_j^* \partial_{p'} \Psi_j - \partial_{p'} \Psi_j^* \partial_{p} \Psi_j \right),
\end{equation}
where the summation on $j$ is implied, and $\mathbf{\Psi} = (\Psi_j)_{j=1,2}$ is the normalized eigenvector of $M$ corresponding to the p-mode, and $p$ and $p'$ are directions in parameter space $\{k_r,L_\ell,S\}$ (we ignore $c_\mathrm{s}$ and $N$ as they are not involved in the degeneracy $\mathbf{p_0}$). We decompose $M$ on the Pauli matrices as
\begin{equation}
    M = \frac{c_\mathrm{s}^2 k_r^2 + L_\ell^2 + N^2 + S^2}{2}I_2 + \mathbf{g}\cdot\bm{\sigma},
\end{equation}
where $\bm{\sigma}$ is the 3-vector of Pauli matrices, and
\begin{equation}
    \mathbf{g}(\mathbf{p}) = \begin{pmatrix}
    -c_\mathrm{s} k_r N \\
    - N S\\
    \frac{N^2 - c_\mathrm{s}^2 k_r^2 - L_\ell^2 - S^2}{2}
    \end{pmatrix},
\end{equation}
with $\mathbf{p} = {}^{\top}(k_r,L_\ell,S)$.
Following classical computations found in \cite{Bernevig2013}, one shows that for the p-band
\begin{equation}
    \frac{1}{2\pi}{F}_{p,p'} = (-1)^2 \frac{1}{4\pi}\frac{\mathbf{g}}{ \left\lVert \mathbf{g}\right\rVert^3}\cdot(\partial_p \mathbf{g} \times \partial_{p'}\mathbf{g}),
\end{equation}
which gives a simple expression for the Chern number
\begin{equation}
    \mathcal{C}^p = - \mathrm{sign}\left(\det(\nabla_\mathbf{p}\mathbf{g})\big|_{\mathbf{p_0}}\right).
\end{equation}
Finally, one has
\begin{eqnarray}
    \mathcal{C}^p &=& - \mathrm{sign} 
    \begin{vmatrix}
    -c_\mathrm{s}N & 0 & 0\\
    0 & 0 & -N\\
    0 & -N & 0
    \end{vmatrix} \\
    &=& - \mathrm{sign}(c_\mathrm{s}N^3)\\
    &=& -1.
\end{eqnarray}
Since the sum of the Chern numbers over the different bands is zero, one obtains directly $\mathcal{C}^g = +1$. The theorem of spectral flow ensures that the number of modes in each band varies by $\left|\mathcal{C}^g \right| = \left|\mathcal{C}^p \right| = 1$ when $S$ changes sign. \cite{Faure2019}. More precisely, when $S(r)$ changes sign from negative to positive values, the p-band of the operator $\mathcal{M}$ loses one mode to the g-band. When $S(r)$ changes sign from positive to negative values, the topological mode transits from the g- towards the p-band \cite{Perrot2019}.\\

\begin{figure}[h!]
	\centering{
	\includegraphics[width=0.45\textwidth]{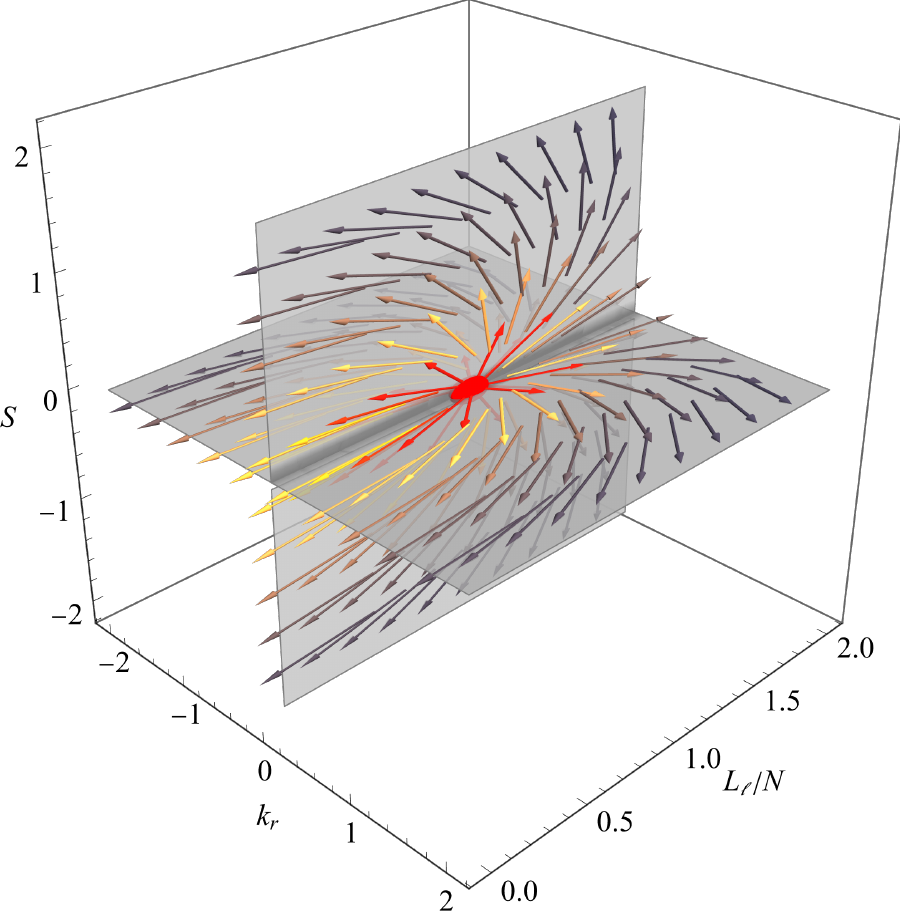}
	\includegraphics[width=0.45\textwidth]{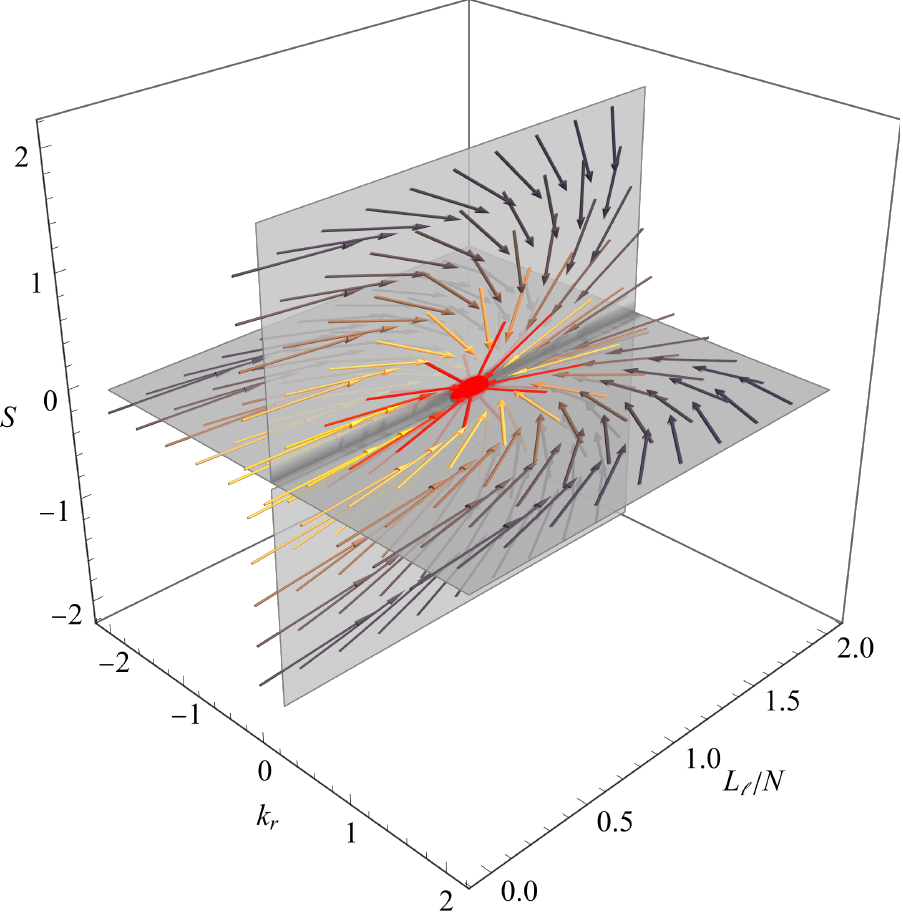}
	}
	\caption{Vector field of the Berry curvature in parameter space $\{k_r,L_\ell,S\}$. Left: $\mathbf{F}^{(1)}$, associated to the g-band (source). Right: $\mathbf{F}^{(2)}$, associated to the p-band (sink). The red dot is the degenerated point at $(0,N,0)$.}
\label{fig:appChern}
\end{figure}

For clarity, detailed expressions of the Berry curvature $\mathbf{F}^{(n)}$ are provided below for both bands, together with representations of the vectors fields in the parameter space (Fig.~\ref{fig:appChern}). One has

\begin{align}
\scriptstyle \mathbf{F}^{(1)} & \scriptstyle =\left(\frac{-4 k L}{\left(\left(k^2+(L-1)^2+S^2\right) \left(k^2+(L+1)^2+S^2\right)\right)^{3/2}},\frac{2 \left(k^2-L^2+S^2+1\right)}{\left(\left(k^2+(L-1)^2+S^2\right) \left(k^2+(L+1)^2+S^2\right)\right)^{3/2}},\frac{-4 L
   S}{\left(\left(k^2+(L-1)^2+S^2\right) \left(k^2+(L+1)^2+S^2\right)\right)^{3/2}}\right), \\
\scriptstyle   \mathbf{F}^{(2)} & \scriptstyle =\left(\frac{4 k L}{\left(\left(k^2+(L-1)^2+S^2\right) \left(k^2+(L+1)^2+S^2\right)\right)^{3/2}},\frac{-2 \left(k^2-L^2+S^2+1\right)}{\left(\left(k^2+(L-1)^2+S^2\right) \left(k^2+(L+1)^2+S^2\right)\right)^{3/2}},\frac{4 L
   S}{\left(\left(k^2+(L-1)^2+S^2\right) \left(k^2+(L+1)^2+S^2\right)\right)^{3/2}}\right).
\end{align}

Poles of the curvature are found to be at the two points $k=S=L_\ell\pm N=0$, as shown on Fig.~\ref{fig:appChern}.

\newpage
%%%%%%%%%%%%%%%%%%%%%%%%%%%%%%
\section{Regularity at the center}
\label{app:CL}
Let us verify that the change of variables Eq.~\eqref{eq:transform} used does not include diverging modes in the spectrum. In the vicinity of the center, one has 

\begin{eqnarray}
    c_\mathrm{s} &\rightarrow& c_\mathrm{s,0},\\
    N &\rightarrow& 0,\\
    S/c_\mathrm{s} &\sim& r^{-1}, \\
    k_{r,0}^2 &\sim& -L_\ell^2/c_\mathrm{s,0}^2,\\
    L_\ell^2/c_\mathrm{s}^2  &\sim& \ell(\ell+1) r^{-2},
\end{eqnarray}
such that Eq.~\eqref{eq:schroWell} becomes
\begin{equation}
       \left(
       \frac{\mathrm{d}^2}{\mathrm{d}r^2}
       -\frac{\ell(\ell+1)}{r^2}
       \right)p = 0 .
\end{equation}
This equation has two solutions, only one of which is regular, which is
\begin{equation}
    p \propto r^\ell.
\end{equation}
We inverse the transform Eq.~\eqref{eq:transform} to obtain the behavior of physical quantities of the perturbation, which are
\begin{eqnarray}
    p' &\propto& r^{\ell-1},\\
    w',v',u' &\propto& r^{\ell-2}.
\end{eqnarray}
As a consequence the radial flux as well as kinetic energy remain finite at the center:
\begin{eqnarray}
    F(r\rightarrow0) &\sim& 4\pi r^2 w' \propto r^\ell,\\
    E_\mathrm{kin} &\equiv& 4\pi \int_0^R \rho \bm{V}^2 r^2 \mathrm{d}r \propto \int_0^R r^{2(\ell-1)},
\end{eqnarray}
which is finite for $\ell>1$. The radial pulsations case $\ell=0$ is left aside, as the topological has zero frequency in this case. The behavior at the other boundary $r=R$ is depend on the given model.

%%%%%%%%%%%%%%%%%%%%%%%%%%%%%%%%%%%%%%%%%%%%%

%%%%%%%%%%%%%%%%%%%%%%%%%%%%%%

%%%%%%%%%%%%%%%%%%%%%%%%%%%%%%
\section{Lamb-like and f-mode}
\label{app:fmode}

The f-mode is defined by Cowling as the stellar mode with zero node in the radial direction \cite{Cowling1941}. Since the topological mode as well as the surface-gravity wave has zero node, we lead numerical experiments to study their coexistence. The Lamb-like wave is present in the spectrum when $S$ changes sign somewhere in the bulk. The surface-gravity wave is present in the spectrum when peculiar boundary conditions are enforced at the surface, namely Poisson's boundary conditions 
\begin{equation}
\partial_t p = \frac{g}{c_\mathrm{s}}w .
\end{equation}
We lead numerical experiments in plane-parallel geometry, with $z$-direction being stratified, and $x$-direction being invariant by translation. We note $z_2$ the top of the medium, and $z_1$ the bottom. The average localisation of a normalized mode $\Psi = (v,w,\Theta,p)$ is the average position of its energy
$$
\bar{z}_\Psi \equiv \int_{z_1}^{z_2} \mathrm{d}z\:z\: \Psi\cdot\Psi^*,
$$
since the sum of kinetic and potential energy of the mode is $\Psi\cdot\Psi^*$.

\begin{figure}[h!]
	\centering{\includegraphics[width=0.6\textwidth]{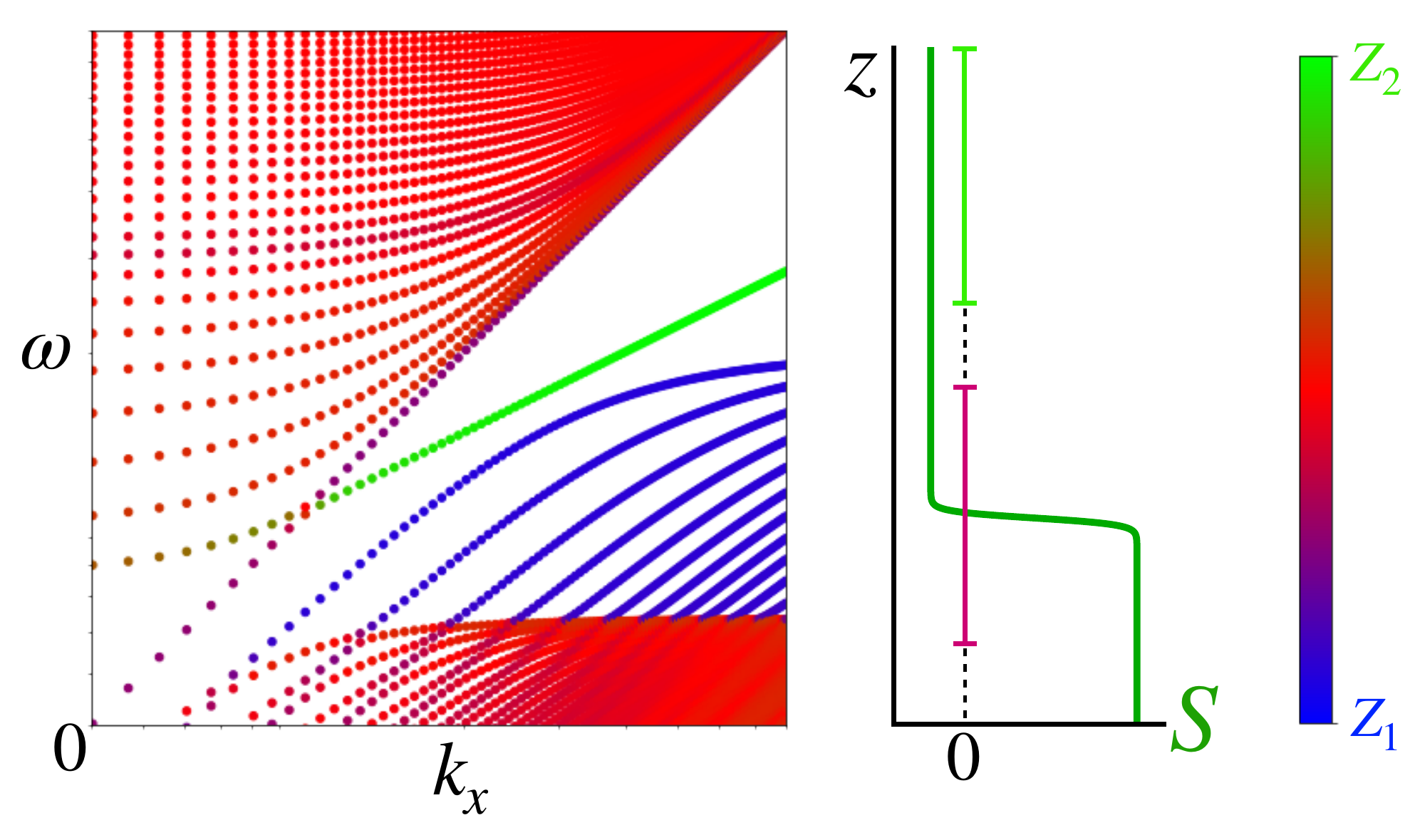}}
	\caption{Spectrum of the waves in a stratified medium with 2D plane-parallel geometry. The buoyancy frequency $N$ is set as a constant. $S$ is set according to the profile displayed on the right panel of the figure. The Lamb-like wave develops around the cancellation point of $S$ (pink interval). The surface-gravity wave is trapped at the top surface (green interval). The two waves do not overlap significantly: the system have two modes with zero nodes.}
\label{fig:f_mode_noHybrid}
\end{figure}

\begin{figure}[h!]
	\centering{\includegraphics[width=0.95\textwidth]{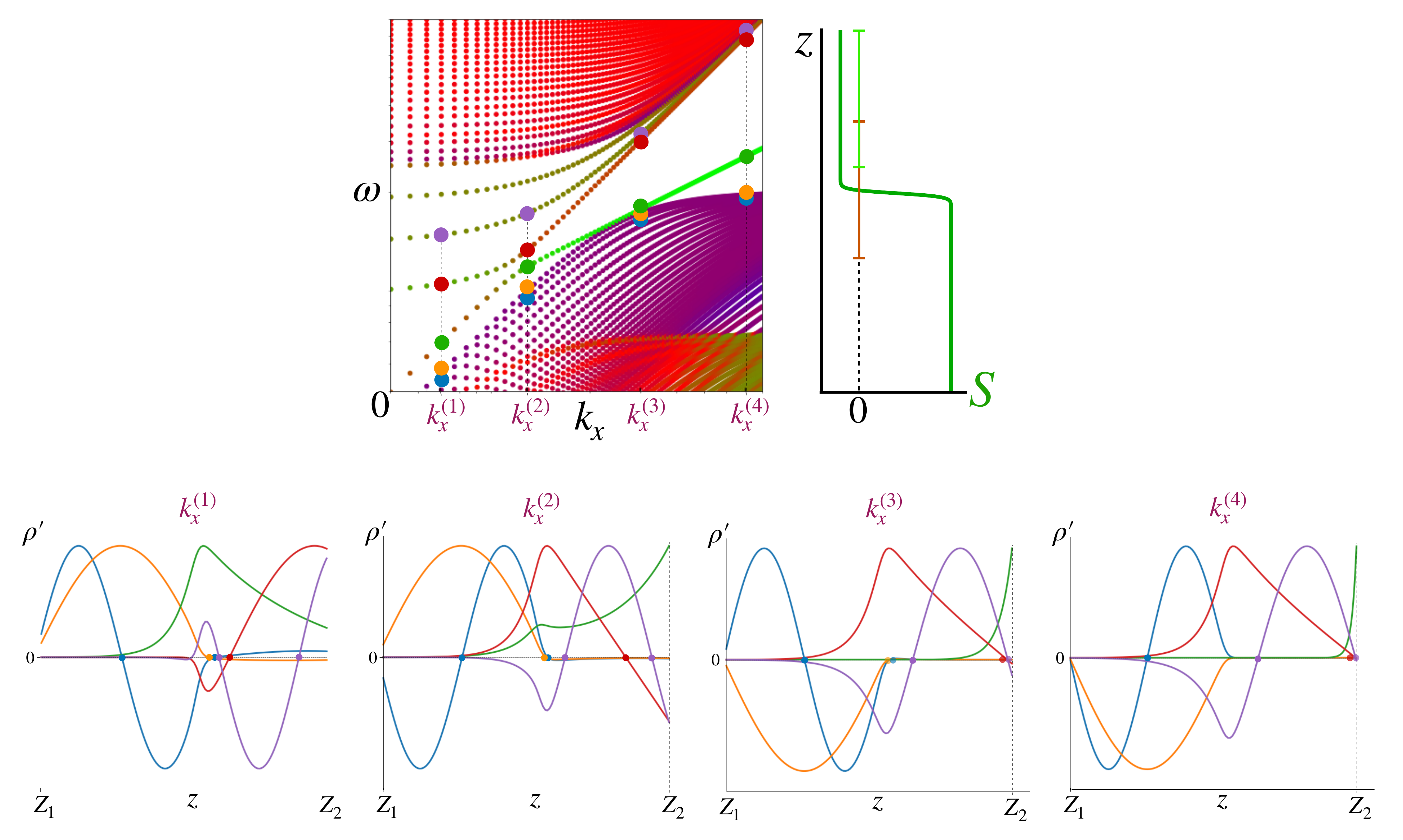}}
	\caption{Spectrum of the same problem as presented in Fig.~\ref{fig:f_mode_noHybrid}, for a profile of $S$ whose zero is closer to the surface $z=z_2$. TheLamb-like and surface-gravity waves overlap and hybrid. This results in one $n=0$ mode: the f-mode.}
\label{fig:f_mode_hybrid}
\end{figure}

In Fig.~\ref{fig:f_mode_noHybrid}, we show that if the Lamb-like is trapped sufficiently far away from the top-surface, it does not hybrid with the surface-gravity wave. In Fig.~\ref{fig:f_mode_hybrid}, we show that if they overlap, they hybrid into a single zero-node mode. We also show the eigenfunctions of the density of the perturbation of a few modes.

\clearpage
\bibliography{biblio}{}
\bibliographystyle{aasjournal}




\end{document}